\documentstyle[aps,prb,multicol,epsf]{revtex}
\begin{document}
\bibliographystyle{prsty}

\title{The importance of being discrete - life always wins on the
surface }
\author{Nadav M. Shnerb,Yoram Louzoun,Eldad Bettelheim,Sorin Solomon}
\address{Department   of Physics, Hebrew   University,
Jerusalem, Israel 91904}

\date{\today}
\maketitle

\def\be{\begin{equation}}
\def\ee{\end{equation}}

\thispagestyle{empty}
\noindent
\begin{abstract}
Many systems in chemistry, biology, finance and social sciences present
emerging
features which are not easy to guess from the elementary interactions of
their
microscopic individual components.

In the past, the macroscopic behavior of such systems was modeled by
assuming
that the collective dynamics of microscopic components can be
effectively described
collectively by equations acting on spatially continuous density
distributions.
It turns out that quite contrary, taking into account the actual
individual/discrete
character of the microscopic components of these systems is crucial for
explaining
their macroscopic behavior.

In fact, we find that in conditions in which the continuum approach
would predict the
extinction of all the population (respectively the vanishing of the
invested capital or of
the concentration of a chemical substance, etc),
the microscopic granularity insures the emergence of macroscopic
localized
sub-populations with collective adaptive properties which allow their
survival and development. In particular it is found that in 2 dimensions "life"
(the localized proliferating phase) always prevails.
\end{abstract}

\begin{multicols}{2}
In addition to physics, an increasing range of sciences:
chemistry, biology, ecology, finance, urban and social planning have
passed
in the last century to quantitative mathematical methods.

Along with the obvious benefits, it turns out that the traditional
differential equations approach has brought some fallacy into their
study.

We present here a very simple generic model which contains proliferating
(and dying)
individuals and we show that in reality it behaves very differently than
its representation
in terms of continuum density distributions:
In conditions in which the continuum equations predict the population
extinction,
the individuals self-organize in spatio-temporally localized adaptive
patches which
insure their survival and development.

This phenomenon admits multiple interpretations in various fields:

-if the individuals are interpreted as interacting molecules, the
resulting
chemical system emerges spatial patches of high density which evolve
adaptively in a way similar with the first self-sustaing systems
which might have anticipated living cells.

- if the individuals are the carriers of specific genotypes represented
in the genetic space, the patches can be identified with species,
which rather than becoming extinct, evolve between various genomes
(locations in the genetic space)  by abandoning regions of
low viability in favor of more viable regions. This adaptive speciation
behavior
emerges in spite of the total randomness we assume for the individuals
motions
in the genetic space (mutations).

- interpreted as financial traders, the individuals develop a "herding"
behavior
in spite of the fact that we do not introduce communication or
interaction between them.
This leads to the flourishing of markets which the continuum analysis
would doom to extinction.

All these phenomena have in common the emergence of large, macroscopic
structures from apparently uniform background \cite{anderson} due to the
amplification of
small, microscopic fluctuations which originate in  the individualized
character of the elementary components of the system.
This mechanism insures in particular that on large enough
2 dimensional surfaces, even if the average growth rate is negative
(due to very large death rate), adaptive structures always emerge and
flourish.

Imagine an area inhabited by a population of eternal agents
A which are spread out uniformly with average  density $n_A$ and move
around randomly,
with diffusion coefficient $D_A$.
Imagine now a race of mortals, B, which are also spread over this area,
with
initial uniform density $n_B(0)$.
The B agents die at a constant rate, $\mu$, ($B
\stackrel{\mu}{\longrightarrow} \emptyset)$
and  proliferate (divide) when they meet the ``catalysator'', A,  with
rate  $\lambda$
($B +A \stackrel{\lambda}{\longrightarrow} B +B +A$).
The B's are diffusive, hopping at the rate $D_B$. What will happen?

The naive lore
based on macroscopic continuity assumptions will predict that A reaches
a spatially
homogeneous distribution, $n_A(x) = n_A$, while the B time variation
${\partial n_{B} \over \partial t}$ is represented by the
linear differential partial differential equation:
\be\label{LPDE}
{\partial n_{B} \over \partial t} = D_B \nabla^2 n_B + (\lambda
n_A-\mu)n_B
\ee
The first term represents the uniformization effect of B diffusion while
the $\mu B$ term represents the fact that a certain fraction of B's die
per unit time.
The crucial term $\lambda n_A n_B$ represents the proliferation of B's
in the presence
of the "life giving" A's.
Note that the equation is linear in $n_B$ and that for initial spatially
uniform
$n_A$ and $n_B$ distributions it has the time exponential solution
\be\label{1}
 n_B (t) = n_B (0) e^{ t (\lambda n_A  - \mu)}.
\ee
In particular Eq. (\ref{1}), predicts that if the
macroscopic proliferation rate
$\lambda n_A $ is lower than the death rate $\mu$,
the B population will uniformly decrease to extinction.

Using microscopic representation techniques\cite{solomon}, one finds
that
populations of discrete  proliferating agents are much more resilient
than
one would first guess based on macroscopic or continuum (PDE) treatment
(Fig. 1).

{\narrowtext
\begin{figure}[h]
\vspace{-0.15in}
\begin{center}
\leavevmode
\epsfxsize=6.0cm
\epsfbox{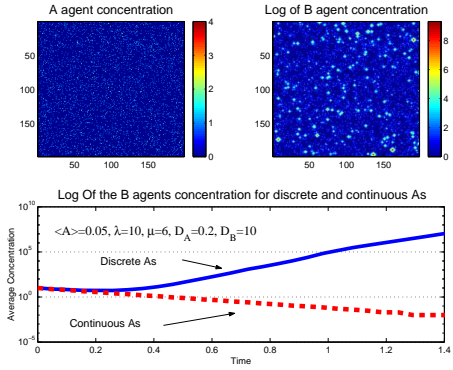} 
\end{center}
\caption{
The lower half of the figure shows the exponential growth of the average
B population as
 a function of time, in the actual simulation (solid blue line) compared
to the
 exponential decrease predicted by the continuum approximation (dashed
red
 line).
$<A>$ is the average
number of A reactants per site. Other symbols are defined in the text.
The
snapshots in the upper half show the spatial configuration of A and B
reactans. B
reactans are seen to be localized in islands (notice that what is
plotted is
the logarithm of B concentrations, thus localization is stronger than
would
first appear).
}
\vspace{-0.26cm}
\end{figure}
}

The study of diffusion limited reactions \cite{dlr} have already shown
in the past
deviations from the continuum theory due to the quantized nature of the
reactants.
In the present case the effect is even more dramatic:
it constitutes the difference between life and death:
the continuum approach predicts extinction while the direct simulation
uncovers the emergence of a thriving, adaptive, developing system (Fig
1).

In order to understand the "source of life" in this system
one has to concentrate on the  microscopic conditions around
the individual A agents, rather then looking at the local average growth
rate
$\lambda n_A - \mu$.

Fig. 2
represents the evolution of the B cloud following a {\it single}
 A agent as it jumps around randomly. The B concentration is shown
to trace the A  as it performs a random walk.
Clearly, the colony does not decay to
extinction ; instead, it seems to trace the A, ``trying'' to keep its
center
of mass at its location. As shown in the inset, each jump of the A is
followed
by a momentary decrease in the height of the B concentration, however,
due to the multiplicative  process there is an overall increase.

{\narrowtext
\begin{figure}[h]
\vspace{-0.15in}
\begin{center}
\leavevmode
\epsfxsize=6.0cm
\epsfbox{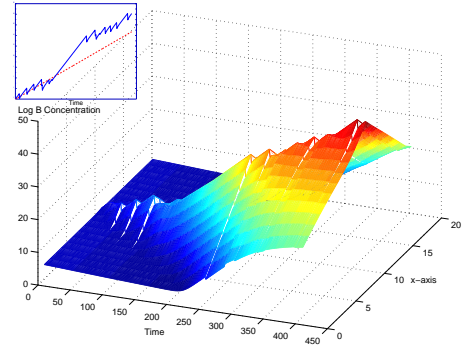} 
\end{center}
\caption{
The profile of a B island as a function of time as it follows the random
motion of an A agent.
The cross-section of the island is taken through the current location of
the A agent.
The inset shows the time evolution of the height of the
B concentration at the point at which A is currently located (solid blue
line).
The B colony is seen to
grow, although the average growth rate over the entire space is negative
($n_A$ is extrenly low since
there is only one in the whole simulation space, thus $\lambda n_A
-\mu\approx
-\mu$).
The dashed red line shows
the and exponential growth with coefficient $\epsilon_0-2dD_A\kappa$
where
$\kappa=Log_e (\lambda/D_B)$, is the slope of the island (this slope is
exhibted in the main graph, and can be derived from a simple approximate
calculation). $\epsilon_0$ is $\lambda -\mu-2dD_B$ derived similarly.
}
\vspace{-0.26cm}
\end{figure}
}

Let us consider first the simplest situation
of a single A agent jumping randomly (with a diffusion coefficent $D_A$)
between the  locations in an infinite d-dimensional space.
In between A jumps the B density at the A location grows exponentially
\cite{nelson}
as $n_B(t) \sim n_B(0) e^{(\lambda -\mu-2 d D_B)t}$. Where $\lambda$,
$\mu$ and
$2dD_B$ stand for the proliferation, death, and the loss due to
diffusion,
respectively. The estimation is made by neglecting the flow of B's from
neighbouring site to the A site, this is justified when the B concentration
in the 
neighbouring sites is much lower than on the A site. In the same limit,
the ratio between the height
of the B density at the A location and
the height of the B density on a neighboring site is easily estimated:
 $\lambda/D_B$.
Consequently, each A jump corresponds to a sudden downwards jump
by a factor of $\lambda/ D_B$ in the height of the B hill . As there are
in
average $2 d D_A$ such jumps per unit time,
the net effect of proliferation, diffusion and death, gives the B
concentration at the
A site as a function of time:
\be\label{3}
n_B (t) = n_B (0) e^{(\lambda -\mu-2dD_B-2dD_A Log_e (\lambda/D_B) )t}
\ee
 The approximation is in good agreement with the
simulation shown in Fig. 2. The slope of the island, on a log scale, is
indeed seen to be $Log_e(\lambda/D_B)$, the the time dependence of the
height of the B island in between A
jumps is indeed given approximately by an exponent, $(\lambda -\mu-2 d
D_B)t$.
Consequently the dashed red line (in the inset) which represents Eq. 3
follows closely the actual growth seen in the
simulation (blue line). The difference between the theory and simulation
is mainly due to
cases where two or more A jumps follow each other rapidly, in this case
the
island's shape does not stabilize before another A jump is made, these
rather
rare events, modify somewhat the actualy result
 \footnote{The above analysis turns
void if $D_B = 0$, where the spatial dimensions of the island
do not grow at all. We do not consider this singular case
in this paper.}.

One may ask what is the situation in the case, when single colonies
are unstable (i.e where the exponent in Eq. 3 is negative).
One possibility is that in such a situation the continuum approximation
is valid and the B concentration decays to zero. Another possibility
is that, although single isolated colonies are unstable, global effects
such as islands
growing, joining and splitting give us back the survival feature. In
particular,
since large colonies are more stable than small colonies, one may expect
the typical size of an ``active'' colony to grow with time. This
behavior
 is demonstrated in Fig. 3 which show the
active clusters in a two dimensional system developing in time.
Evidently, the small clusters either decay or merge into larger and
larger clusters.

{\narrowtext
\begin{figure}[h]
\vspace{-0.15in}
\begin{center}
\leavevmode
\epsfxsize=8.0cm
\epsfbox{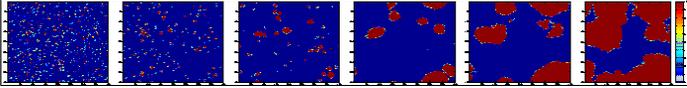} 
\end{center}
\caption{
The spatial distribution of B's for different times
(earlier times to the left), where concentrations
greater than 10 are colored red and concentrations below 10 are color
coded.}
\vspace{-0.26cm}
\end{figure}
}

The arguments and computer experiments mentioned  above
show convincingly that the individuals based life is much
more resilient than a hypothetical life density spread continuously
across spatial
regions.

These conclusions should suffice to induce professionals in biology,
finance and social
sciences to consider seriously the regime in which their systems are
naively non-viable
(decay to extinction) when viewed macroscopically but perfectly viable
in reality
(and when simulated correctly at the microscopic individual level
\cite{nick}).

In particular, most of the species in nature could be in this regime:
negative
naive
average macroscopic growth rate but actual survival and proliferation.
Similarly, markets which might look unappealing when averaging over the
various
investing possibilities might prove lucrative enough (at least for the
lucky
investors which hit profit opportunities A) as to maintain them in the
competitive range.
In fact this line of thought might provide an explanation to the
emergence of
life from
the random chemistry of its component molecules in spite of the formal
extreme
improbability of the event.
Equally it might explain the paradoxes in finance between the efficient
market hypothesis
(absence of systematic profit opportunities in equilibrium markets) and
the actual
profits which investors extract daily from the market.

In order to obtain a more rigorous bound on the parameter range in which
life overcomes the gloom prognosis of the  macroscopic
analysis
we used the renormalization group (RG) analysis which indicates that on
large enough surfaces, life always wins.
For higher dimensions, the dominion of life still extends
to arbitrary low $n_A$ densities, but a minimal finite $\lambda$ value
is required.

In RG, the collective behavior  of the
system is identified by integrating out the small length scale, short
time fluctuations, leaving us with an effective theory  for the
large scale objects. Here, these are the large,
stable islands shown in Fig. 3. The new, effective theory
is
characterized by  renormalized coupling constants, i.e., modified
numerical values of the effective rates (growth rate, dearth rate,
hopping
etc.)  on large length scale. The process of  decimating small
fluctuations  is then iterated
again and again, giving us flow line which reflect the evolution of  the
effective values of the coupling constants as one integrates larger and
larger  scales $l$.

\footnote{
The details of this RG analysis, which involves the
presentation
of the exact Master equation of the process as a field integral
and the  $\epsilon$-expansion around the critical dimension $d_c=2$
are out of the scope of this report and will be presented elsewhere.}

Fig. 4
shows the flow lines of m ($m=\mu- \lambda n_A$) and $\lambda$ due to
the iteration of the
decimation
process\footnote{ $D \equiv D_A + D_B$ is the effective diffusion
constant}. The flow is given by the equations:
\begin{eqnarray}
{dm \over dl} = 2 m-{\lambda^2 n_A  \over 2 \pi D} \nonumber \\
{d \lambda  \over dl} = \lambda [ 2-d +{\lambda   \over 2 \pi D}]
\end{eqnarray}
 For  $d \leq 2$ we see that for large length and time
scales (that is, after many iterations of the decimation process),
$\lambda$
grows without limit while $m$ eventally becomes negative. This implies
that on the large scale, the system actually behaves as if $\lambda n_A
> \mu$,
and life always wins.

\vspace{1in}
{\narrowtext
\begin{figure}[h]
\vspace{0in}
\begin{center}
\leavevmode
\epsfxsize=5cm
\epsfbox{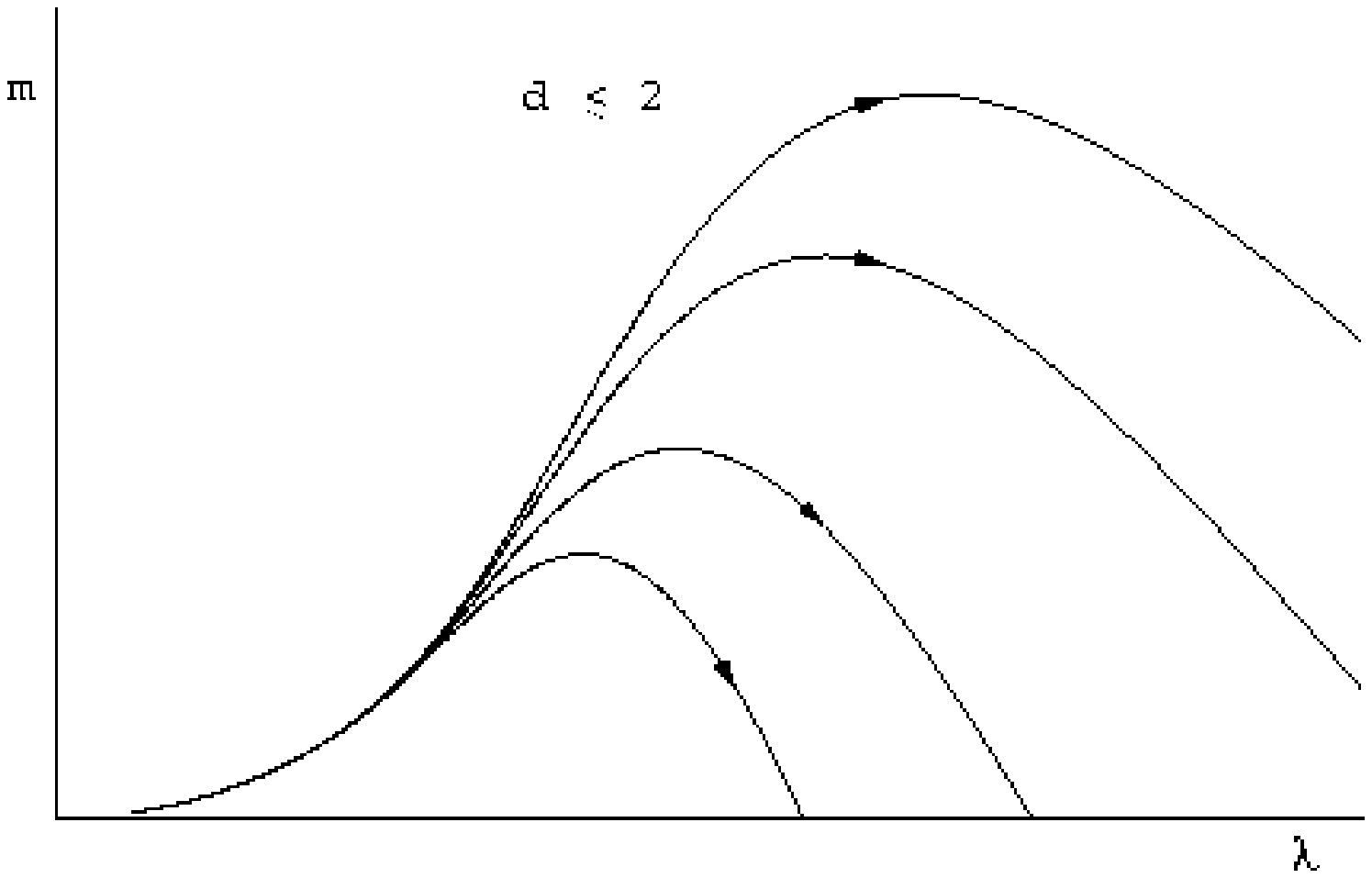} 
\end{center}
\vspace{0.3in}
\begin{center}
\leavevmode
\epsfxsize=5cm
\epsfbox{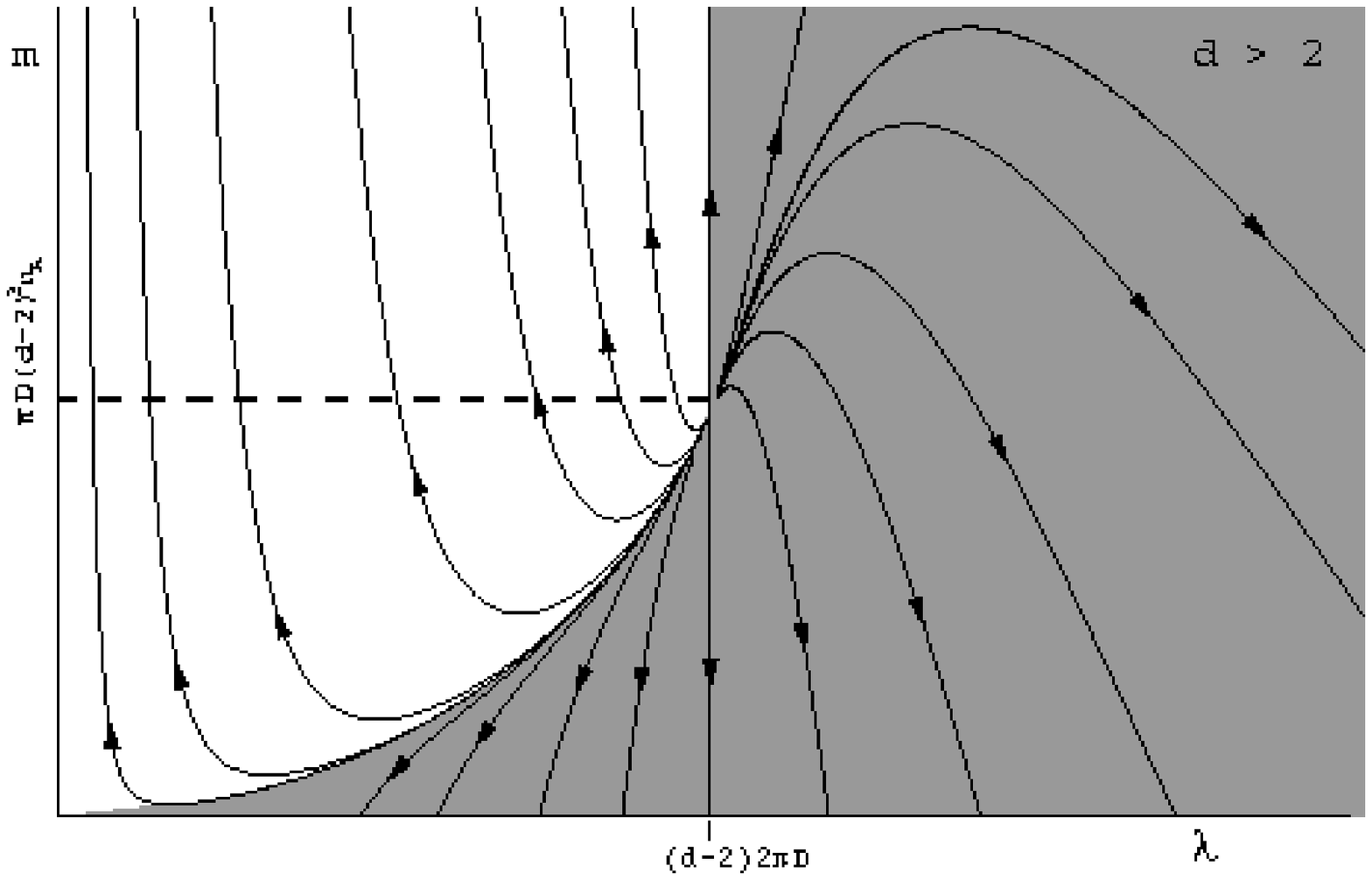} 
\end{center}
 \caption{
Lower panel shows flow lines for $d > 2$. Shaded region flows to negative mass
("life"). Upper panel shows flow lines for $d \leq  2$, the whole parameter space flows
to negative mass.
}
\vspace{-0.26cm}
\end{figure}
}

In higher dimensions ($ d > 2 $) Fig. 4{\bf a} indicates a
dynamical phase
 transition where for part of the parameter space the system flows to
negative
$m$ (life) and for another part the system flows to positive $m$
(death).

 It should be noted that the flow  portrayed in Fig.
4 is associated with larger and larger length scales:
For a finite system, the flows should be truncated and
the size of the system may be crucial:
simulations with parameters identical to that of Fig. 3,
lead to extinction
when carried out on a system size 4 times smaller.

In conclusion, our results suggest that the dimensionality of the system
and its size
are crucial features for its capability to emerge and  sustain life.
This may
explain the fact that most of the ecological systems are two
dimensional.
Reinterpreting in  the genome space, the present results provide the
conceptual link
between the atomized structure of the life building blocks and the
explosive Darwinian tandem, noise $+$ proliferation.

\end{multicols}
\end{document}